# Advanced simulation framework for AC/MTDC power systems

**Aleksandra Lekić**
**Delft University of Technology**
**Netherlands**
**a.lekic@tudelft.nl**

**Azadeh Kermansaravi***
**The Hague University of Applied Sciences**
**Netherlands**
**azadeh.kermansaravi@ieee.org**

**Haixiao Li**
**Delft University of Technology**
**Netherlands**
**h.li-16@tudelft.nl**

**Yasel Quintero Lares**
**Delft University of Technology**
**Netherlands**
**y.i.quinterolares@tudelft.nl**

**Saif Alsarayreh**
**Delft University of Technology**
**Netherlands**
**salsarayreh@tudelft.nl**

**Robert Dimitrovski**
**TenneT TSO GmbH**
**Germany**
**R.Dimitrovski@tudelft.nl**

**SUMMARY**

Alternating current (AC)/multi-terminal direct current (MTDC) hybrid power systems (HPSs) play a crucial role in enabling long-distance power transmission and flexible interconnections between AC grids. However, the challenges that HPSs encountered are numerous, with stability and harmonic issues being particularly prominent. Traditional electromagnetic transient (EMT) tools have struggled to accommodate small-signal stability problems and the potential issues of the optimal interactions among converters. To address this gap, HARMONY ("**HARMON**ic stabilit**Y** assessment of PE-penetrated power systems") has been developed for the advanced simulation and analysis of interconnected AC/MTDC HPSs as a comprehensive mathematical framework based on C++ programming language. The primary goals of Harmony are to provide faster and trusted stability analyses, and address the analytical difficulties associated with converter control dynamics, converter-driven stability, and interoperability in HPSs. This framework is intended to be open source, therefore broadening collaboration for researchers, and to contribute to the community of power systems engineers.

In this paper, we demonstrate two core functionalities featured in HARMONY, that are optimal power flow (OPF) and harmonic stability analyses (HAS). The underlying analysis models and computational methodologies for both functionalities are presented in detail to help future readers and users gain a clear understanding of mathematical fundamentals of HARMONY. Furthermore, we introduce the integrated framework of OPF and HAS designed in HARMONY, along with representative printed analysis results, to demonstrate the appealing capabilities of HARMONY.

**KEYWORDS**

Multi-terminal DC, harmonic stability analysis, optimal power flow, open source

# 1 Introduction

With the increasing global emphasis on clean energy, renewable energy sources (RESs), like wind and solar, are expanding to large-scale. To facilitate their efficient conversion, transmission, and accommodation, voltage source converters (VSC) based alternating current (AC)/multi-terminal direct current (MTDC) hybrid power grids (HPSs) are rapidly emerging, as a key composition for future power systems. However, the integration of VSCs are bringing new requirements for the system operation and the stability when considering their interactions.

Compared with traditional AC system stability, converter controllers interact with the surrounding power system and with each other over a wide frequency range, leading to so-called *harmonic* or *electromagnetic stability* issues [1] . Such stability can be assessed in the Laplace or frequency domain using state-space or impedance-based methods [2]. Although EMT tools are widely used for detailed time-domain analysis, as they can explicitly capture switching dynamics and high-frequency interactions of power electronic converters, their high computational burden makes them unsuitable for fast stability screening across operating points. Consequently, impedance-based stability analysis has emerged as an efficient and promising approach for VSC-based AC/MTDC hybrid power systems [1, 3–6]. However, despite its conceptual advantages, dedicated software support for impedance-based small-signal stability analysis remains limited, particularly for systems with detailed converter and network models.

It is worth noting that two fresh tools for impedance/admittance characterization and small-signal stability assessment have been released, namely, Z-tool [7], a Python-based implementation, and PowerImpedanceACDC [8], developed in Julia. These tools mark important steps toward open-source frequency-domain stability analysis. However, as the scale and complexity of AC/MTDC HPSs continue to grow, the computational requirements for real-time analysis are expected to increase dramatically. To pursue extreme computational capability, we aim to develop a harmonic stability analysis (HSA) tool based on C++ programming language. Compared with Python and Julia implementations, the C++ based architecture is able to achieve higher execution speed and fine-grained memory control. C++ supports seamless integration with high-performance numerical libraries, which together contribute to improved scalability and robustness. Unlike interpreted or JIT-compiled environments, C++ allows the development of fully standalone executables or libraries. As a result, the developed tool can be directly executed without additional environment configuration, simplifying deployment in engineering practice. In addition, the developed tool is envisioned to serve as a comprehensive platform that also incorporates fruitful functionalities such as optimal power flow (OPF) computation. This enables a dual-perspective assessment of AC/MTDC HPSs, where both operational stability and optimal performance can be evaluated within a unified framework.

Driven by these emerging needs for high computational performance, integrated analysis, and open accessibility, the C++-based HARMONY tool has been conceived and developed. It includes detailed component modelling, providing HSA and OPF for AC/MTDC HPSs.

# 1 HPS Harmonic Stability Analysis

The HVDC system includes harmonics in the grid. These harmonics entail serious stability issues and a real source of problems. HSA uses all conventional frequency-domain stability methods, including eigenvalues, participation factors, Bode, and Nyquist plots. The main advantage of this approach is that everything is integrated into a single toolbox. In this section, key steps of HPS are presened. It should be noted that the algorithm described below is applicable to balanced systems.

## 1.1 Adjusted modified nodal analysis approach

To effectively compute the equivalent impedance for a given circuit, the adjusted modified nodal analysis (MNA) approach is utilized [3], which is applied with following steps:

***Step 1.*** **-**Determine the input and output of the circuit where you want to determine the equivalent impedance, see Figure 1.

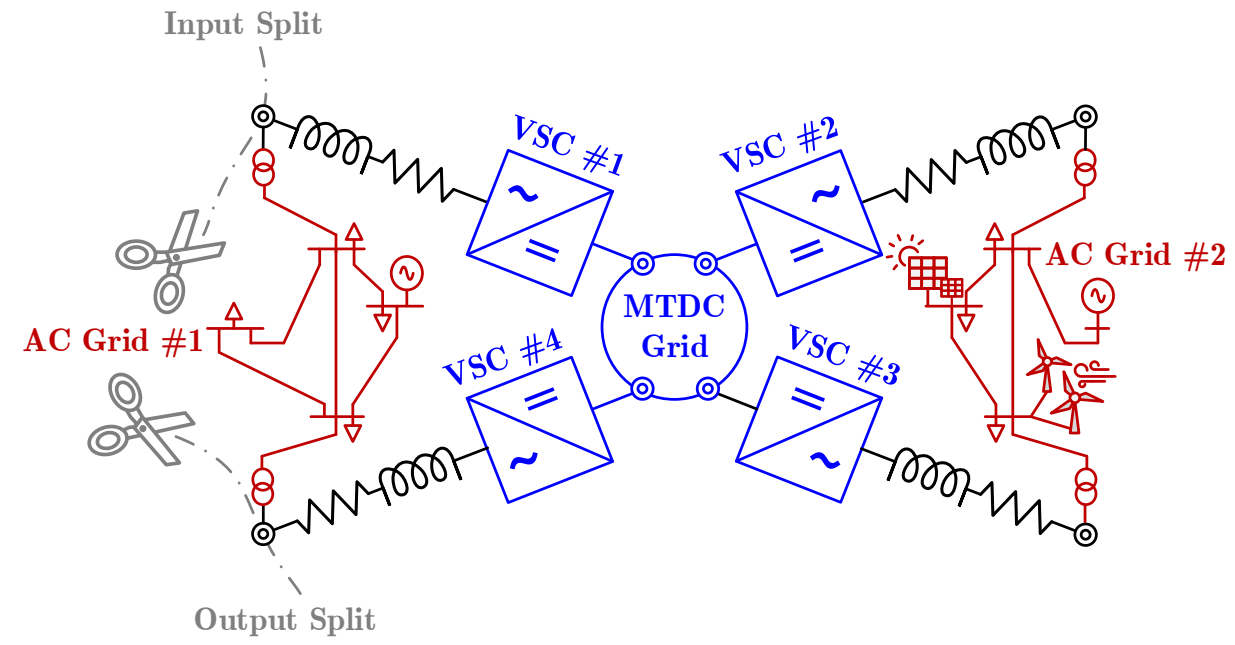

***Figure 1*** ***–Schematic diagram regarding circuit cut***

***Step 2.*** **-**Assign input and output current as depicted in Figure 2.

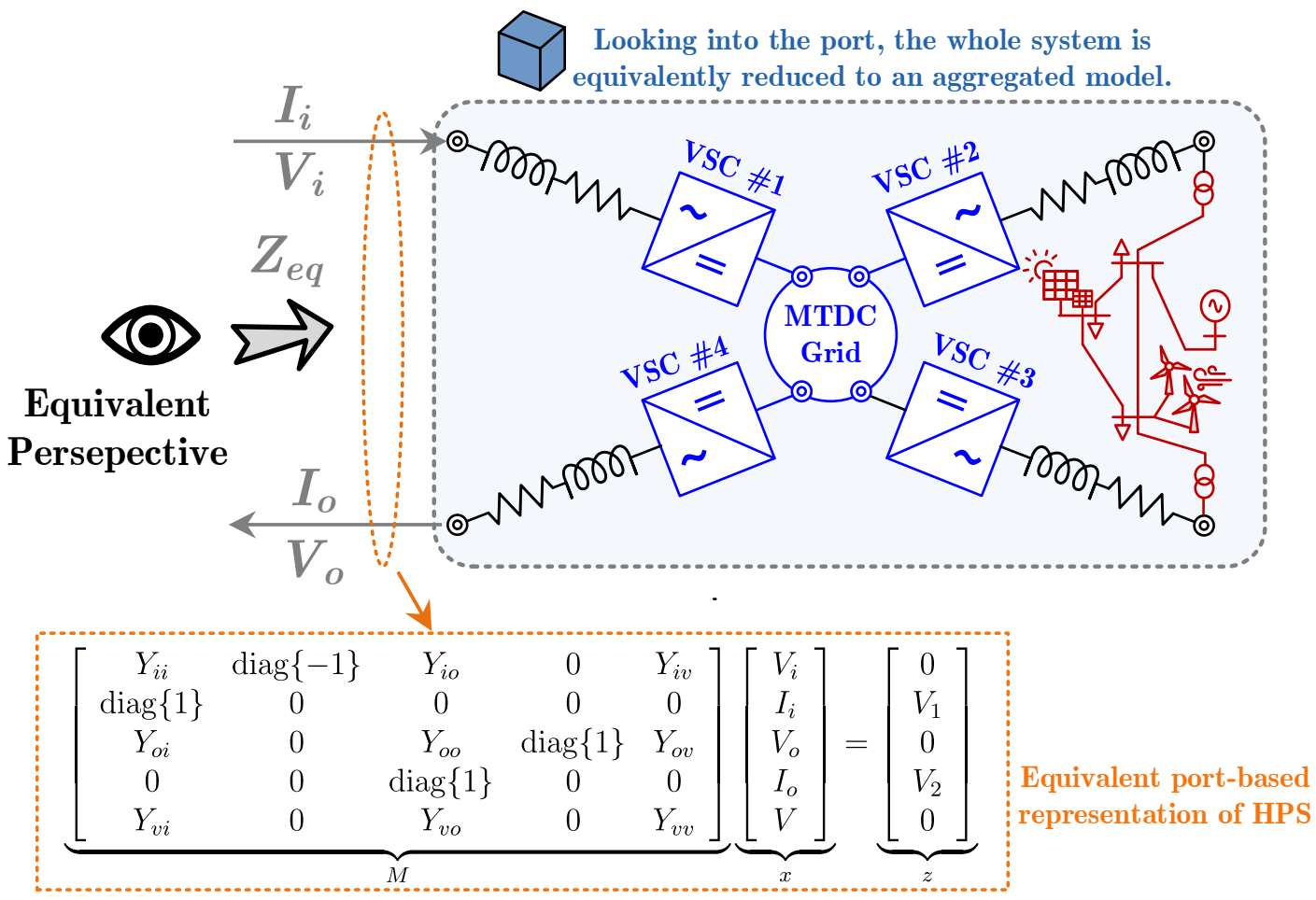

$$\underbrace{\begin{bmatrix} Y_{ii} & \mathrm{diag}\{-1\} & Y_{io} & 0 & Y_{iv} \\ \mathrm{diag}\{1\} & 0 & 0 & 0 & 0 \\ Y_{oi} & 0 & Y_{oo} & \mathrm{diag}\{1\} & Y_{ov} \\ 0 & 0 & \mathrm{diag}\{1\} & 0 & 0 \\ Y_{vi} & 0 & Y_{vo} & 0 & Y_{vv} \end{bmatrix}}_{M} \underbrace{\begin{bmatrix} V_i \\ I_i \\ V_o \\ I_o \\ V \end{bmatrix}}_{x} = \underbrace{\begin{bmatrix} 0 \\ V_1 \\ 0 \\ V_2 \\ 0 \end{bmatrix}}_{z}$$

***Figure 2*** ***- Schematic diagram regarding equivalent impedance determination***

***Step 3.*** **-**Build MNA by adding all components except the ones on the direct connection to the input and output pins. Those components are considered as non-existent as seen in Figure 2. Furthermore, the following sorting of variables should be applied: $[V_i, I_i, V_o, I_o, V]$, where $V_i$ and $I_i$ present voltages and currents at the input like depicted in Figure 2. $V_o$ and $I_o$ are output voltages and currents, and $V$ are other voltage buses/nodes that are not input and output ones.

***Step 4.*** **-**The formulated equations will be follows:

$$\underbrace{\begin{bmatrix} Y_{ii} & \text{diag}\{-1\} & Y_{io} & 0 & Y_{iv} \\ \text{diag}\{1\} & 0 & 0 & 0 & 0 \\ Y_{oi} & 0 & Y_{oo} & \text{diag}\{1\} & Y_{ov} \\ 0 & 0 & \text{diag}\{1\} & 0 & 0 \\ Y_{vi} & 0 & Y_{vo} & 0 & Y_{vv} \end{bmatrix}}_{M} \underbrace{\begin{bmatrix} V_i \\ I_i \\ V_o \\ I_o \\ V \end{bmatrix}}_{x} = \underbrace{\begin{bmatrix} 0 \\ V_1 \\ 0 \\ V_2 \\ 0 \end{bmatrix}}_{z}, \quad (1)$$

where $Y_{xy}$ are Y parameters for the specified block of nodes $x$ and $y$. Values $V_1$ and $V_2$ are symbolic vectors of values assigned to input and output nodes. With 0 is denoted zero matrix. For simplicity, the dimensions are skipped in the previous equation.

***Step 5.*** **-**The previous (1) is solved using the reduced row echelon form [4] on the concatenated matrices $\mathbf{M}$ and $z$. This will solve the system of equations by enforcing the values for variables $x$ to diagonal 1 values, and on the position of $z$ column would be expected solution.

***Step 6.*** **-**By taking values for $V_i$ and $I_i$ and dividing them, we obtain the equivalent impedance per phase. For estimating the equivalent admittance parameters, the subnetwork is defined together with its outputs. Outputs represent start buses, while end buses do not exist.

Therefore, the previous algorithm is changed to:

$$\underbrace{\begin{bmatrix} Y_{ii} & \text{diag}\{-1\} & Y_{iv} \\ \text{diag}\{1\} & 0 & 0 \\ Y_{vi} & 0 & Y_{vv} \end{bmatrix}}_{M} \underbrace{\begin{bmatrix} V_i \\ I_i \\ V \end{bmatrix}}_{x} = \underbrace{\begin{bmatrix} 0 \\ V_1 \\ 0 \end{bmatrix}}_{z} \quad (2)$$

and Y parameters are obtained by exchanging one by one of the inputs with 1 and others set to 0 in $V_1$, and determining admittances correspondingly.

## 1.2 Stability assessment procedure

The stability assessment procedure follows the steps:

***Step 1.*** **-**Divide components per AC and DC areas, and converter group.

***Step 2.*** **-**Estimate equivalent admittance parameters of the subnetwork between outputs (connections to converters) for each AC and DC area. Please note that the AC side need to be transformed to the DQ frame [5], [6]

$$\mathbf{Y}_{dq} = \frac{1}{6}\left\{\mathbf{a}\mathbf{Y}\big(j(\omega-\omega_0)\big)\mathbf{a}^H + \frac{1}{6}\mathbf{a}^*\mathbf{Y}\mathbf{Y}\big(j(\omega+\omega_0)\big)\mathbf{a}^H\right\}_{dq}, \quad (3)$$

The derivation follows the corrected theorem from [7]. Namely, starting from Fourier transformation of the Park transform $P_{\omega_0}(j\omega)$ and $P_{\omega_0}^{-1}(j\omega)$ as:

$$P_{\omega_0}(j\omega) = \frac{2\pi}{3}(\mathbf{a}\delta(\omega-\omega_0) + \mathbf{a}^*\delta(\omega+\omega_0) + \mathbf{c}\delta(\omega)), \quad (4)$$

and

$$P_{\omega_0}^{-1}(j\omega) = \pi(\mathbf{a}^T\delta(\omega-\omega_0) + \mathbf{a}^H\delta(\omega+\omega_0) + 2\mathbf{c}^T\delta(\omega)), \quad (5)$$

where $\mathbf{a}^*$ is elementwise complex conjugate of $\mathbf{a}$, and $\mathbf{a}^H = (\mathbf{a}^*)^T$ is Hermitian transpose of $\mathbf{a}$. Furthermore, we have that

$$\mathbf{a}=\begin{bmatrix}1 & \exp(j\varphi) & \exp(2j\varphi)\\ j & j\exp(j\varphi) & j\exp(2j\varphi)\\ 0 & 0 & 0\end{bmatrix},\ \mathbf{c}=\begin{bmatrix}0 & 0 & 0\\ 0 & 0 & 0\\ 1 & 1 & 1\end{bmatrix}. \tag{6}$$

Starting from relationship $I_{abc}(j\omega)=Y(j\omega)V_{abc}(j\omega)$, when applying the Park transformation, we get:

$$I_{dqz}(j\omega)=\frac{1}{2\pi}P_{\omega_0}(j\omega)\otimes\left(Y(j\omega)\frac{1}{2\pi}P_{\omega_0}^{-1}(j\omega)V_{dqz}(j\omega)\right), \tag{7}$$

which gives:

$$\mathbf{I}_{dqz}(j\omega)=\frac{1}{6}\mathbf{aY}\big(j(\omega-\omega_0)\big)\mathbf{a}^T\mathbf{V}_{dqz}\big(j(\omega-2\omega_0)\big)$$
$$+\frac{1}{6}\mathbf{aY}\big(j(\omega-\omega_0)\big)\mathbf{a}^H\mathbf{V}_{dqz}(j\omega)+\frac{1}{3}\mathbf{aY}\big(j(\omega-\omega_0)\big)\mathbf{c}^T\mathbf{V}_{dqz}\big(j(\omega-\omega_0)\big)$$
$$+\frac{1}{6}\mathbf{a}^*\mathbf{Y}\big(j(\omega+\omega_0)\big)\mathbf{a}^T\mathbf{V}_{dqz}(j\omega)+\frac{1}{6}\mathbf{aY}\big(j(\omega+\omega_0)\big)\mathbf{a}^H\mathbf{V}_{dqz}\big(j(\omega+2\omega_0)\big)$$
$$+\frac{1}{6}\mathbf{cY}(j\omega)\mathbf{a}^T\mathbf{V}_{dqz}\big(j(\omega-\omega_0)\big)+\frac{1}{6}\mathbf{aY}(j\omega)\mathbf{a}^H\mathbf{V}_{dqz}\big(j(\omega+\omega_0)\big)$$
$$\frac{1}{3}\mathbf{a}^*\mathbf{Y}\big(j(\omega+\omega_0)\big)\mathbf{c}^T\mathbf{V}_{dqz}\big(j(\omega+\omega_0)\big)+\frac{1}{3}\mathbf{cY}(j\omega)\mathbf{c}^T\mathbf{V}_{dqz}(j\omega). \tag{8}$$

Since everything that multiplies **c** gives zero when we take dq-frame part (first 2 rows and columns), and since we have that $\mathbf{a} *\times (\cdot)\times \mathbf{a}^{\boldsymbol{T}}$ and $\mathbf{a} *\times (\cdot)\times \mathbf{a}^{\boldsymbol{H}}$ is zero for dq-variables (condition is that upper and lower diagonal values are equal module, but different sign and diagonal values are the same), the expression becomes:

$$\mathbf{I}_{dqz}(j\omega)=\frac{1}{6}\left(\mathbf{aY}\big(j(\omega+\omega_0)\big)\mathbf{a}^H+\mathbf{a}^*\mathbf{Y}\big(j(\omega+\omega_0)\big)\mathbf{a}^T\right)V_{dq}(j\omega). \tag{9}$$

***Step 3.*** **-**Use formulas for converter cross-connections on DC side. Namely, for the DC admittance visible inside the converter, we use the already known AC side admittance denoted as $Y_{eq,AC}$. Then we have that:

$$i_{dq}^{\Delta}=Y_{dq,conv}v_{dq}+a_{2\times1,conv}v_{dc}=Y_{eq,AC}v_{dq}^{\Delta}, \tag{10}$$

$$v_{dq}^{\Delta}=(Y_{eq,AC}-Y_{dq,conv})^{-1}a_{2\times1,conv}v_{dc}, \tag{11}$$

$$i_{dc}=\underbrace{\left(b_{1\times2,conv}\left(Y_{eq,AC}-Y_{dq,conv}\right)^{-1}a_{2\times1,conv}+Y_{dc,conv}\right)}_{Y_{eq,conv}}v_{dc}, \tag{12}$$

which is the value used in further calculation. This case is applicable for every converter that is not considered for the local stability check, and for the DC cutting side of the converter, for the estimation of local stability. Please note that the calculated admittance parameters $Y_{eq,AC}$ in this case are also the input in the AC network when looking from the AC side of the converter.

***Step 4.*** **-**Find closing impedance on DC side. Here, we consider equivalent admittance parameters of the DC of the converter, which is examined for stability check, and that bus connection (with converter) we consider an input. The other DC network outputs are considered outputs closed with "closing impedance" determined in the previous step.

The closing impedance will look as $Y_L = \mathrm{diag}\{Y_{eq,conv1}, \cdots, Y_{eq,convN}\}$, for $N$ being the number of converters which are not examined for stability (i.e. one less than the total number of converters in the system, not considering the converters included as a part of RES), and $Z_{eq,conv} = Y_{eq,conv}^{-1}$.

Then we have that the admittance parameters can be divided into $Y_{11}$, $Y_{12}$, $Y_{13}$, and $Y_{22}$ after identifying the input and outputs, and have that:

$$I_{in} = Y_{11}V_{in} + Y_{12}V_{out}, I_{out} = Y_{21}V_{in} + Y_{22}V_{out}, V_{out} = -Z_L I_{out}, V_{out} = -Z_L I_{out}, \tag{13}$$

and finally, we get that:

$$I_{in} = \underbrace{Y_{11} + Y_{12}(Y_L + Y_{22})Y_{21}}_{Y_{eq}} V_{in}. \tag{14}$$

and finally, we get that:

***Step 5.*** **-**Couple AC and DC sides of the converter of interest. Here we identify two cases:

- ✧ If the system is cut on the DC side, then we will use the formula for the DC side cross-coupling (7), and the admittance on its AC side calculated in ***Step 1***, and calculate the transfer as: $TF(s) = Y_{eq,conv}Y_{eq,DC}^{-1}$.
- ✧ If the system is cut on the AC side, we use the admittance calculated in ***Step 4***, estimate the admittance visible at the converter's AC side as follows. To estimate the AC admittance, we follow the same procedure, and we have

$$i_{dc} = b_{1\times2,conv}v_{dq}^{\Delta} + Y_{dc,conv}v_{dc} = Y_{eq,DC}v_{dc}, \tag{15}$$

$$v_{dc} = \frac{b_{1\times2,conv}v_{dq}^{\Delta}}{Y_{eq,DC} - Y_{dc,conv}}, \tag{16}$$

$$i_{dq}^{\Delta} = \underbrace{\left(Y_{dq,conv} + \frac{a_{2\times1,conv}b_{1\times2,conv}}{Y_{eq,DC} - Y_{dc,conv}}\right)}_{Y_{eq,conv}} v_{dc}. \tag{17}$$

Finally, calculate the transfer function with the AC admittance calculated in ***Step 1*** as $F(s) = Y_{eq,conv}Y_{eq,AC}^{-1}$.

## 2 AC/DC OPF for HPSs

AC/DC OPF determines the steady-state operating points of HPSs and directly affects system impedance characteristics and harmonic interactions. This section presents an SOCP-based formulation implemented in HARMONY, which is extensible to future developments involving switching actions and integer variables.

- **Sets, Indexes, Scripts:** $\mathcal{N}/\mathcal{E}$ is the node/branch set; $\overline{(\cdot)}/\underline{(\cdot)}$ is the upper/lower limit; $(\cdot)^{AC}/(\cdot)^{MTDC}/(\cdot)^{VSC}$ is the parameter or variable that are related to AC, MTDC and VSC.

**AC grid branch**

$$p_{ij}^{AC} = g_{ij}^{AC}(c_{ii}^{AC} - c_{ij}^{AC}) + b_{ij}^{AC}s_{ij}^{AC}, \quad \forall i \in \mathcal{N}^{AC}, \forall(ij) \in \mathcal{E}^{AC} \tag{18}$$

$$q_{ij}^{AC} = b_{ij}^{AC}(c_{ii}^{AC} - c_{ij}^{AC}) + g_{ij}^{AC}s_{ij}^{AC}, \quad \forall i \in \mathcal{N}^{AC}, \forall(ij) \in \mathcal{E}^{AC} \tag{19}$$

**AC grid bus**

$$p_i^{AC} = \sum_{ij} p_{ij}^{AC} + c_{ii}^{AC} G_{i,sh}^{AC}, \quad \forall i \in \mathcal{N}^{AC}, \forall(ij) \in \mathcal{E}^{AC} \tag{20}$$
$$q_i^{AC} = \sum_{ij} q_{ij}^{AC} - c_{ii}^{AC} B_{i,sh}^{AC}, \quad \forall i \in \mathcal{N}^{AC}, \forall(ij) \in \mathcal{E}^{AC} \tag{21}$$
$$p_i^{AC} = p_{i,gen}^{AC} + p_{i,res}^{AC} - p_{i,load}^{AC} - p_{i,A2V}^{AC}, \quad \forall i \in \mathcal{N}^{AC} \tag{22}$$
$$q_i^{AC} = q_{i,gen}^{AC} + q_{i,res}^{AC} - q_{i,load}^{AC} - q_{i,A2V}^{AC}, \quad \forall i \in \mathcal{N}^{AC} \tag{23}$$
$$(\underline{v}_i^{AC})^2 \le c_{ii}^{AC} \le (\overline{v}_i^{AC})^2, \quad \forall i \in \mathcal{N}^{AC} \tag{24}$$

**Generator and RES**

$$\underline{p}_{i,gen}^{AC} \le p_{i,gen}^{AC} \le \overline{p}_{i,gen}^{AC}, \quad \forall i \in \mathcal{N}^{AC} \tag{25}$$
$$\underline{q}_{i,gen}^{AC} \le q_{i,gen}^{AC} \le \overline{q}_{i,gen}^{AC}, \quad \forall i \in \mathcal{N}^{AC} \tag{26}$$
$$0 \le p_{i,res}^{AC} \le \overline{p}_{i,res}^{AC}, \quad \forall i \in \mathcal{N}^{AC} \tag{27}$$
$$-\overline{s}_{i,res}^{AC} \le \cos\left(\frac{k}{N}\pi\right) p_{i,res}^{AC} + \sin\left(\frac{k}{N}\pi\right) q_{i,res}^{AC} \le \overline{s}_{i,res}^{AC}, \quad \forall i \in \mathcal{N}^{AC} \tag{28}$$

**SOCP relaxation**

$$(c_{ij}^{AC})^2 + (s_{ij}^{AC})^2 \le c_{ii}^{AC} c_{jj}^{AC}, \quad \forall i \in \mathcal{N}^{AC}, \forall(ij) \in \mathcal{E}^{AC} \tag{29}$$

- **AC grid constraints:** (18) and (19) represents the AC branch flow. $p_{ij}^{AC}/q_{ij}^{AC}$ is the AC branch real/reactive power. $g_{ij}^{AC}/b_{ij}^{AC}$ is the AC branch conductance/susceptance. $c_{ii}^{AC}, c_{ij}^{AC}, s_{ij}^{AC}$ are the new variables related to AC nodal voltage. More specifically, there is symmetry that $c_{ij}^{AC} = c_{ji}^{AC}, s_{ij}^{AC} = -s_{ji}^{AC}$ ; (20) and (21) represent the AC bus power injection, according to the nodal power balance principle. $G_{i,sh}^{AC}/B_{i,sh}^{AC}$ is the real/imaginary parts of the shunt admittance of the AC bus; (22) and (23) specify the contribution of the AC bus power injection. $p_{i,gen}^{AC}/q_{i,gen}^{AC}$ is the real/reactive power production of the generator at the AC bus. $p_{i,res}^{AC}/q_{i,res}^{AC}$ is the real/reactive power production of the RES at the AC bus. $p_{i,load}^{AC}/q_{i,load}^{AC}$ is the real/reactive power consumption at the AC bus. $p_{i,A2V}^{AC}/q_{i,A2V}^{AC}$ is the real/reactive power transmission form the AC grid to the VSC; (24) regulates the squared AC bus voltage limits. (25) and (26) regulate the generator power output limits. (27) regulate the active power limits of the RES; (28) depicts the use of polygonal approximation to represent the originally circular feasible power region [15]; (29) forms a SOCP relaxation of an intrinsic equation that $(c_{ij}^{AC})^2 + (s_{ij}^{AC})^2 \le c_{ii}^{AC} c_{jj}^{AC}$.

**DC grid branch**

$$p_{jh}^{DC} + p_{hj}^{DC} = r_{jh}^{DC} l_{jh}^{DC}, \quad \forall i \in \mathcal{N}^{DC}, \forall(jh) \in \mathcal{E}^{DC} \tag{30}$$
$$u_j^{DC} - u_h^{DC} = r_{jh}^{DC}(p_{jh}^{DC} - p_{hj}^{DC}), \quad \forall j,h \in \mathcal{N}^{DC}, \forall(jh) \in \mathcal{E}^{DC} \tag{31}$$

**DC grid bus**

$$p_j^{DC} = \rho^{DC} \sum_{jh} p_{jh}^{DC}, \quad \forall j \in \mathcal{N}^{DC}, \forall(jh) \in \mathcal{E}^{DC} \tag{32}$$
$$(\underline{v}_j^{DC})^2 \le u_j^{DC} \le (\overline{v}_j^{DC})^2, \quad \forall j \in \mathcal{N}^{DC}, \tag{33}$$

**DC side control (optional)**

$$p_j^{DC} = p_{j,ref}^{DC}, \quad \forall j \in \mathcal{N}^{DC}, \tag{34}$$
$$u_j^{DC} = (v_{j,ref}^{DC})^2, \quad \forall j \in \mathcal{N}^{DC}, \tag{35}$$
$$p_j^{DC} = p_{j,ref}^{DC} - \frac{1}{k_{j,droop}^{DC}}\left(\frac{1}{2} + \frac{1}{2}u_j^{DC} - v_{j,ref}^{DC}\right), \quad \forall j \in \mathcal{N}^{DC}, \tag{36}$$

**SOCP relaxation**

$$(p_{jh}^{DC})^2 \le l_{jh}^{DC} u_j^{DC}, \quad \forall i \in \mathcal{N}^{DC}, \forall(ij) \in \mathcal{E}^{DC} \tag{37}$$

- **DC grid constraints:** (30) represents the power loss relationship on the DC branch. $p_{jh}^{DC}/l_{jh}^{DC}$ is the power/current following along the DC branch. $r_{jh}^{DC}$ is the DC branch resistance; (31) represents the branch voltage drop on the DC branch. $u_j^{DC}, u_h^{DC}$ are the squared voltages at the two terminal buses. (32) represents the DC bus power. $\rho^{DC}$ means polarity of the DC network. $\rho^{DC} = 1$ represents the monopolar and $\rho^{DC} = 2$ represents the bipolar. (33) represents the squared DC bus voltage limits. (34), (35), and (36) formulate the DC side control. If the control parameters on the VSC DC side are preset, then the control constraints on the DC side should be activated. (34) represents the DC constant power control and $p_{j,ref}^{DC}$ is the preset DC power reference. (35) represents the DC constant voltage control and $v_{j,ref}^{DC}$ is the preset DC power reference. (35) represents the DC droop control. $k_{j,droop}^{DC}$ is the droop slope. The approximation of $\frac{1}{2}+\frac{1}{2}u_j^{DC} \approx v_j^{DC}$ is obtained via the first-order Taylor expansion of $\sqrt{u_j^{DC}}$, with around $u_j^{DC} = 1\,\mathrm{p.u.}$. (37)(37)(37) forms a SOCP relaxation of an intrinsic equation that $(p_{jh}^{DC})^2 = l_{jh}^{DC} u_j^{DC}$.

| | | |
|---|---|---|
| **VSC branch** | | |
| $p_{sf}^{VSC} = c_{sf}^{VSC} g_{sf}^{VSC} - c_{sf}^{VSC} g_{sf}^{VSC} + s_{sf}^{VSC} b_{sf}^{VSC},$ | $(sf) \in \mathcal{E}^{VSC}$ | (38) |
| $q_{sf}^{VSC} = -c_{sf}^{VSC} b_{sf}^{VSC} + c_{sf}^{VSC} b_{sf}^{VSC} + s_{sf}^{VSC} g_{sf}^{VSC},$ | $(sf) \in \mathcal{E}^{VSC}$ | (39) |
| $p_{cf}^{VSC} = c_{cf}^{VSC} g_{cf}^{VSC} - c_{cf}^{VSC} g_{cf}^{VSC} + s_{cf}^{VSC} b_{cf}^{VSC},$ | $(cf) \in \mathcal{E}^{VSC}$ | (40) |
| $q_{cf}^{VSC} = -c_{cf}^{VSC} b_{cf}^{VSC} + c_{cf}^{VSC} b_{cf}^{VSC} + s_{cf}^{VSC} g_{cf}^{VSC},$ | $(cf) \in \mathcal{E}^{VSC}$ | (41) |
| **VSC bus** | | |
| $p_s^{VSC} = p_{sf}^{VSC} = p_{i,A2V}^{AC},$ | $s \in \mathcal{N}^{VSC}, i \in \mathcal{N}^{AC}$ | (42) |
| $q_s^{VSC} = q_{sf}^{VSC} = q_{i,A2V}^{AC},$ | $s \in \mathcal{N}^{VSC}, i \in \mathcal{N}^{AC}$ | (43) |
| $p_c^{VSC} = p_{cf}^{VSC},$ | $s \in \mathcal{N}^{VSC}, (sf) \in \mathcal{E}^{VSC}$ | (44) |
| $q_c^{VSC} = q_{cf}^{VSC},$ | $s \in \mathcal{N}^{VSC}, (sf) \in \mathcal{E}^{VSC}$ | (45) |
| $q_f^{VSC} = -c_{ff}^{VSC} b_f^{VSC},$ | $f \in \mathcal{N}^{VSC}$ | (46) |
| **VSC current** | | |
| $0 \leq I_c^{VSC} \leq \bar{I}_c^{VSC},$ | $c \in \mathcal{N}^{VSC}$ | (47) |
| $0 \leq l_c^{VSC} \leq \left(\bar{I}_c^{VSC}\right)^2$ | $c \in \mathcal{N}^{VSC}$ | (48) |
| **VSC power loss** | | |
| $p_c^{VSC} + p_{loss}^{VSC} + p_j^{DC} = 0,$ | $c \in \mathcal{N}^{VSC}, j \in \mathcal{N}^{DC},$ | (49) |
| $p_{loss}^{VSC} = a_{loss}^{VSC} + b_{loss}^{VSC} I_c^{VSC} + c_{loss}^{VSC} l_c^{VSC},$ | $\forall j \in \mathcal{N}^{DC},$ | (50) |
| **VSC AC-side control (optional)** | | |
| $c_{ss}^{VSC} = (v_{s,ref}^{VSC})^2,$ | $s \in \mathcal{N}^{VSC}$ | (51) |
| $q_s^{VSC} = q_{s,ref}^{VSC},$ | $s \in \mathcal{N}^{VSC}$ | (52) |
| **SOCP relaxation** | | |
| $(I_c^{VSC})^2 \leq l_c^{VSC},$ | $c \in \mathcal{N}^{VSC}$ | (53) |
| $(p_c^{VSC})^2 + (q_c^{VSC})^2 \leq c_{cc}^{VSC} l_c^{VSC}$ | $c \in \mathcal{N}^{VSC}$ | (54) |

- **VSC constraints:** (38) and (39) represents the branch flow of VSC AC side. More specifically, as illustrated in Fig. 3, it refers to the flow $p_{sf}^{VSC}, q_{sf}^{VSC}$ that are from the PCC bus to the filter point. $g_{sf}^{AC}/b_{sf}^{AC}$ is the corresponding branch conductance/susceptance. Also, (40) and (41) represents the branch flow $p_{cf}^{VSC}, q_{cf}^{VSC}$ from the AC terminal to the filter point. $g_{sf}^{AC}/b_{sf}^{AC}$ is the corresponding branch conductance/susceptance; (42) and (43) refers to the power injection $p_s^{VSC}, q_s^{VSC}$ at the PCC bus; (44) and (45) refers to the power injection $p_f^{VSC}, q_f^{VSC}$ at the AC terminal; (46) refers to the reactive power injection at the filter point. For the latest multilevel topology, the filter is not required and $b_f^{VSC} = 0$; Moreover, $c_{ss}^{VSC}, c_{cc}^{VSC}, c_{ff}^{VSC} c_{fs}^{VSC}, s_{sf}^{VSC}, c_{cf}^{VSC}, c_{fc}^{VSC}$ are the new variables related to VSC nodal voltage. (47) and (48) regulate the bound of the current $I_c^{VSC}$ and the squared current $l_c^{VSC}$; (49) indicates that converter loss $p_{loss}^{VSC}$ generated during power conversion. More specifically, $p_{loss}^{VSC}$ is a quadratic function of $I_c^{VSC}$, $a_{loss}^{VSC}, b_{loss}^{VSC}, c_{loss}^{VSC}$ are the loss coefficients.

(51) and (52) respectively denotes the U-constant and Q-constant control options at the PCC bus. $v_{s,ref}^{VSC}/q_{s,ref}^{VSC}$ is the voltage/reactive power reference for U-constant/Q-constant control. (53) and (54) form the SOCP relaxation. Before relaxation, the equality constraint should be $(I_c^{VSC})^2 = l_c^{VSC}$ and $(p_c^{VSC})^2 + (q_c^{VSC})^2 \leq c_{cc}^{VSC} l_c^{VSC}$.

| Goal: | $\min \sum_{i\in\mathcal{N}^{AC}} c_{i,2}^{AC}(p_{i,gen}^{AC})^2 + c_{i,1}^{AC}\, p_{i,gen}^{AC} + c_{i,0}^{AC},$ | (55) |
|---|---|---|

- **Optimization goal:** (55) represents the optimization goal that is minimizing power generation costs. $c_{i,0}^{AC}/c_{i,1}^{AC}/c_{i,2}^{AC}$ is the constant/linear/quadratic generation cost coefficient.

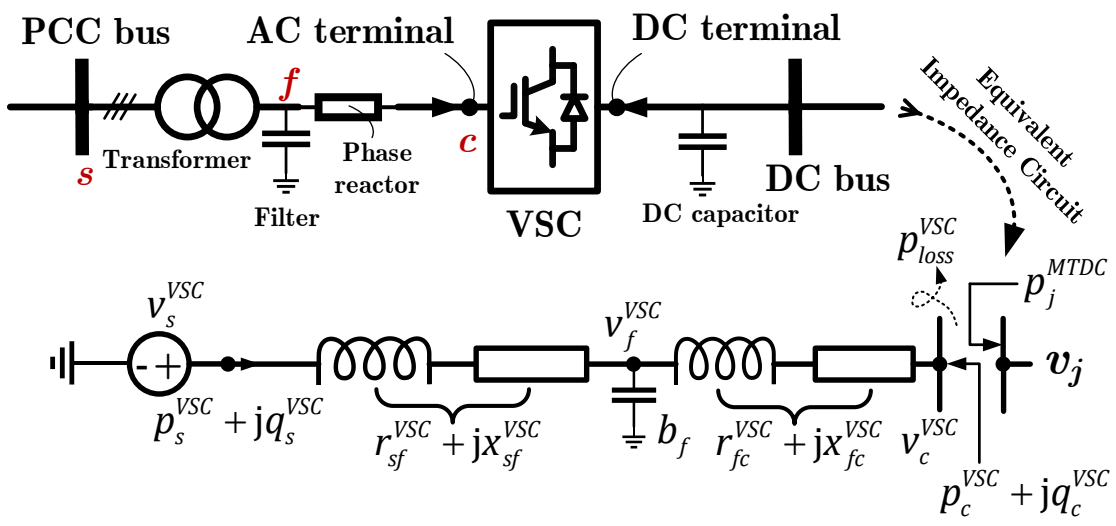

***Figure 3 – Equivalent circuit model of VSC***

We may notice that the bus voltage phase is eliminated after SOCP relaxation. However, the voltage phase is important in harmonic stability analysis of HPSs. Therefore, it is necessary to recover the system-wide voltage phase from the obtained AC/DC OPF results. More details regarding this part can be found in our user Manual [16].

# Implementation of HARMONY

In this section, an overview of HARMONY implementation is presented, including the involved key implement steps, plot support and the printed information of the analysis results.

## 2.1 Key implement steps

Figure 4 illustrates the general flowchart of HARMONY for stability assessment. The whole procedure can be roughly divided into three key implement steps.

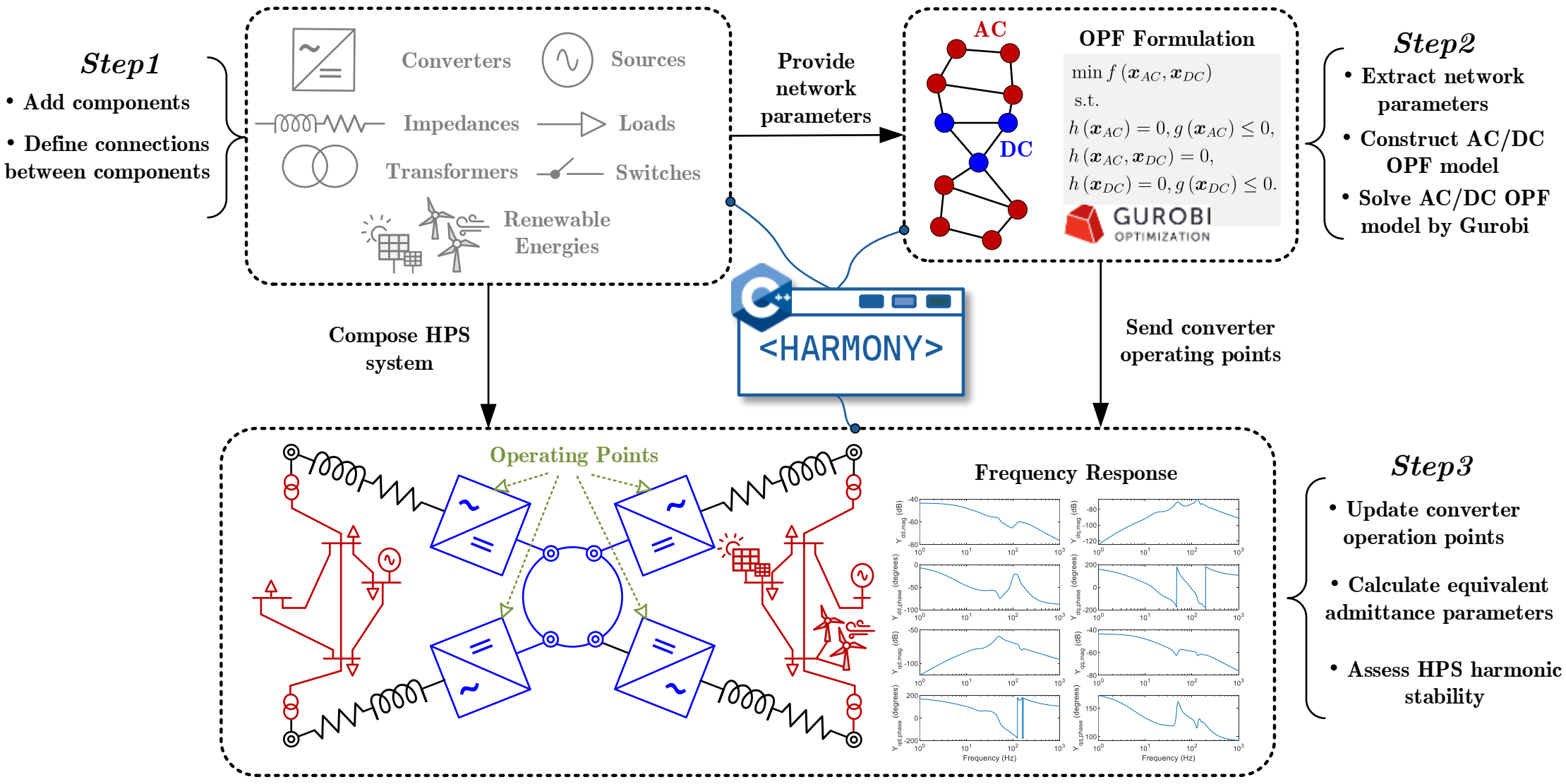

*Figure 4 – General summary of HARMONY implementation*

***Step 1.*** -We need to add the components that form the HPS system. HARMONY provides component class including converters, impedances, loads, transformers, sources, switches, and renewable energies. Parameters must be specified when defining the components, as they are required for both OPF and HAS. After adding all components, a complete HPS system can be constructed by matching the pins at both ends of each component.

***Step 2.*** -By phrasing the formed HPS system, the essential parameters required for OPF can be obtained. Furthermore, a series of constraints related to the AC network, DC network, and converter stations are added, ultimately forming the AC/DC OPF model. HARMONY employs Gurobi[1] to solve the optimization model and operating points of VSCs are produced.

***Step 3.*** -Based on the AC/DC OPF results, the operating points of the VSCs are updated. Then, by computing the equivalent impedance of the HPS, HAS of the system is performed. The specific assessment procedure has been outlined in Section 2.3.

## 2.2 Functionalities for plotting support

Harmony enables fruitful functionalities for plotting support with help of the open-source poltting tool matplot++[2], which are depicted in Figure 5 and detailed as below.

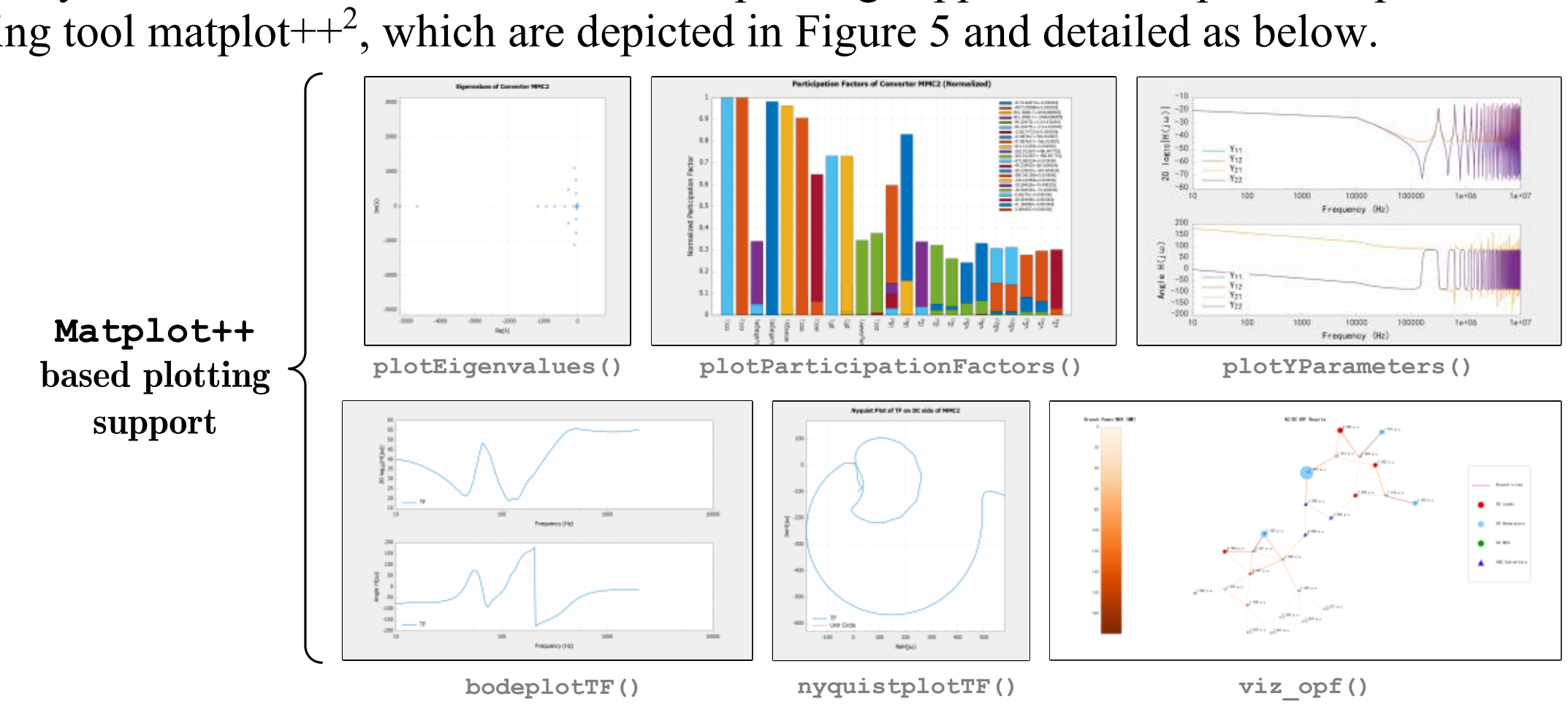

*Figure 5 – Examples of plotting supports*

- **`plotEigenvalues():`** plots eigenvalues of the nonlinear element/component, provide a visual indication of its small-signal stability characteristics around the operating point.
- **`plotParticipationFactors():`** plots participation factors of the nonlinear element/component, helps identify dominant dynamic components.
- **`plotYParameters():`** plots the Y-parameters of the nonlinear element/component, characteres its frequency-domain admittance behavior and facilitating impedance-based stability analysis.
- **`bodeplotTF():`** plots the Bode diagram of the feedback transfer function $Y_{conv}Z_{eq}$ for the desired cut in the AC/HVDC power system. The cut must be next to some of the converters at the AC or DC location.

[1] https://www.gurobi.com/

[2] https://alandefreitas.github.io/matplotplusplus/

- **`nyquistplotTF():`** plots the Nyquist diagram of the feedback transfer function $Y_{conv}Z_{eq}$ for the desired cut in the AC/HVDC power system. The cut must be next to some of the converters at the AC or DC location.
- **`viz_opf():`** visualizes the AC/DC OPF results, provide a graphical overview of network states and power flow distributions.

## 2.3 Access the results

Harmony provides printed information on terminal to access the results. As depicted in Figure 6, when an end-to-end HPS case study is implemented, the corresponding OPF and HSA results will be printed. The OPF output summarizes the steady-state operating conditions of the AC/DC system, including key variables such as bus voltages, power injections, and converter operating points. Based on these operating points, the HSA output then reports the convergence process of the harmonic equilibrium solver for each MMC. Together, these outputs provide a direct link between steady-state optimal operation and frequency-domain stability assessment.

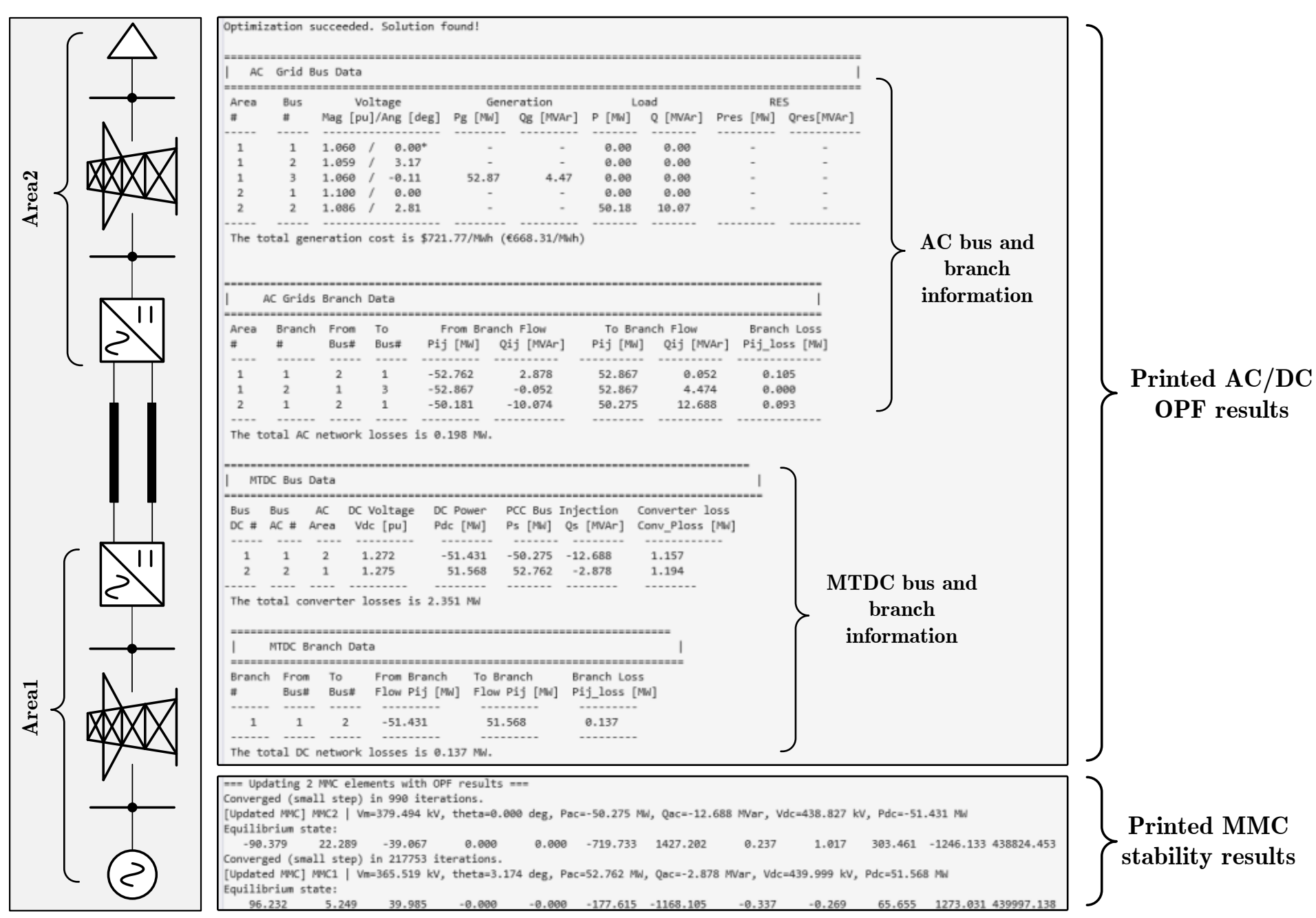

*Figure 6 – Illustration regarding accessing the results in HARMONY*

# 3 Conclusion

HARMONY, an C++-based simulation and analysis tool for HPS, has been introduced. In the current version, this tool has achieved linking between HAS and AD/DC OPF, enabling harmonic analysis under optimized steady-state operating conditions. In future work, HARMONY will be further extended by incorporating dynamic phasor–based component modeling and simulation capabilities. This enhancement will allow a more accurate representation of frequency-coupled dynamics and transient interactions, making the analysis more closely aligned with HPSs. he algorithm currently works for balanced systems. However due to the current implementation of the graph search algorithm, the AC grid connected to two separate terminals should not be interconnected. These limitations will be resolved in the future.